\documentclass[fleqn,11pt]{article}
\usepackage{amsmath,amsfonts,amssymb}

\usepackage{geometry}
\newtheorem{theorem}{Theorem}
\newtheorem{lemma}[theorem]{Lemma}

\newcommand{\sectionnew}[1]{
\setcounter{theorem}{0}
\setcounter{lemma}{0}}

\newcommand{\Tr}{\rm{Tr}}

\begin{document}
\title{\bf A kinetic equation for spin polarized Fermi systems.}
\author{Leif Arkeryd\thanks{Mathematical Sciences, S-41296 Gothenburg, Sweden.}}

\date{}

\maketitle

{\noindent \bf Abstract.}\hspace{0.1in}
  This paper considers a kinetic Boltzmann equation,  having a general type of collision kernel and modelling spin-dependent Fermi gases at low temperatures. The distribution functions have values in the space of positive hermitean $2\times2$ complex matrices. Global existence of weak solutions is proved in $L^1\cap L^{\infty}$ for the initial value problem of this Boltzmann equation in a periodic box.

\footnotetext[2]{Key words; Fermi system with spin, Boltzmann spin kinetics, low temperature kinetics.}

\section{ Introduction.}

The experimental study of spin polarized Fermi gases at low temperatures and their kinetic modelling is well established in physics, an early mathematical physics text in the area being [S]. The first experiments concerned very dilute solutions of $^3He$ in superfluid $^4He$ with - in comparison with classical Boltzmann gases - interesting new
properites such as spin waves (see [NTLCL]). The experimentalists later turned to other set-ups, in particular laser-trapped low temperature gases (see [JR]).\\
Before turning to the mathematical modelling of such phenomena, we recall some basics about the Pauli spin matrices
\begin{eqnarray*}
\sigma_1=\left(\begin{array}{ll}0 & 1\\1 & 0\end{array}\right),\quad \quad
\sigma_2=\left(\begin{array}{ll}0 & i\\-i & 0\end{array}\right),\quad \quad
\sigma_3=\left(\begin{array}{ll}1 & 0\\0 & -1\end{array}\right).
\end{eqnarray*}
Denoting by $[\sigma_i,\sigma_j]$ the commutator $\sigma_i\sigma_j-\sigma_j\sigma_i$, the Pauli matrices satisfy
\begin{eqnarray}
\quad\quad
[\sigma_1,\sigma_2]=2i\sigma_3,\quad
[\sigma_2,\sigma_3]=2i\sigma_1,\quad
[\sigma_3,\sigma_1]=2i\sigma_2,\quad
{\rm and}\quad  [\sigma_i,\sigma_i]=0\quad{\rm for}\quad i=1,2,3.\quad
\end{eqnarray}
With
\begin{eqnarray*}
\sigma=(\sigma_1,\sigma_2,\sigma_3)
\end{eqnarray*}
the Pauli spin vector, (1.1)
is equivalent to $\sigma \times \sigma =2i\sigma$. \\
Let ${\mathcal M}_2(\mathbb C)$ denote the space of $2\times2$ complex matrices with ${\mathcal H}_2(\mathbb C)$ the subspace of hermitean matrices. ${\mathcal H}_2(\mathbb C)$ is linearly isomorphic to $\mathbb R ^4$ using the decomposition $\rho=A_cI+A_s\cdot \sigma$ and identifying $\rho\in{\mathcal H}_2(\mathbb C)$ with $(A_c,A_s)\in\mathbb R^4$.
For $\mathcal{A,B}\in{\mathcal M}_2(\mathbb C)$ the  contracted product of $\mathcal{A}$ and $\mathcal{B}$ is  $\mathcal{A:B}=\sum A_{ij}\bar{B}_{ij}$.
The contracted product of two Pauli matrices $\sigma_i$ and $\sigma_j$ gives
${\sigma}_i:{\sigma}_j=2\delta_{ij}$, where $\delta_{ij}$ is Kronecker's delta. More generally for two vectors
$v=(v_1,v_2,v_3)$ and $w=(w_1,w_2,w_3)$, it holds
 $(v\cdot \sigma):(w\cdot \sigma)=2v\cdot w$.
With $I$ the identity matrix, $(v\cdot \sigma)(w\cdot\sigma)=v\cdot w I+i(v\times w)\cdot\sigma$, which implies $[v\cdot\sigma,w\cdot \sigma]
=2i(v\times w)\cdot\sigma.$ For $\rho^{\dagger}$ the conjugate transpose of $\rho$, it holds $\rho: \rho^{\dagger}=\Tr \rho\rho^{\dagger}=:\parallel \rho\parallel^2$.
\newpage
\hspace{1cm}\\
A dilute spin polarized gas with spin $\frac{1}{2}$, can be modelled kinetically by a distribution function $\rho(t,x,p)$ with values in ${\mathcal H}_2(\mathbb C)$,  which is  the
Wigner transform of the one-atom density operator for the system. The domain of $\rho(t,x,p)$ is positive time $t$, $p\in \mathbb R^3$, and for simplicity in this paper $x\in \mathbb{T}^3$, i.e. periodic 3-dimensional position space with period one.
The number density of particles is given by
$f:= \Tr(\rho(t,x,p))$,
and the magnetization of particles by the vector
$\bar{\sigma}(t,x,p):= Tr(\sigma\rho(t,x,p))$.
It follows that $\rho=\frac{1}{2}(fI+\bar{\sigma}\sigma)$.
In the fermion case it is assumed that $0\leq f\leq 2$, $\min((2-f)^2,f^2)\geq \bar{\sigma}\cdot\bar{\sigma}$. That condition is equivalent to the hermitean matrices $\rho$ and $I-\rho$ having non-negative eigenvalues, which will be denoted $\rho\geq0$, $I-\rho\geq0$.\\
A common model in mathematical physics considers the evolution
\begin{eqnarray}
\frac{\partial f}{\partial t}+\bigtriangledown _p \epsilon_p\cdot \bigtriangledown_x f-\bigtriangledown_x\epsilon_p\bigtriangledown_p \cdot  f+\sum\Big[\frac{\partial h_p}{\partial p_i}\cdot \frac{\partial \bar{\sigma}}{\partial x_i} - \frac{\partial h_p}{\partial x_i}\cdot \frac{\partial \bar{\sigma}}{\partial p_i}\Big] =Q_n ,  \nonumber     \\
\frac{\partial \bar{\sigma}}{\partial t}+\sum_i\Big[ \frac{\partial \epsilon _p}{\partial p_i}\frac{\partial \bar{\sigma}}{\partial x_i}-
 \frac{\partial \epsilon _p}{\partial x_i}\frac{\partial \bar{\sigma}}{\partial p_i}+ \frac{\partial f }{\partial x_i}\frac{\partial h_p}{\partial p_i}- \frac{\partial f_p}{\partial p_i}\frac{\partial h_p}{\partial x_i}\Big]-2(h_p\times \bar{\sigma})=Q_m
\end{eqnarray}
(cf [JM]). Here $Q_n$ is the number density part and $Q_m$ the magnetization part of the collision operator. The energy matrix is split into
\begin{eqnarray*}
\epsilon_p=\frac{p^2}{2m}+\int dp'\{V(0)-\frac{1}{2}V(|p-p'|)\}f(t,x,p'),    \\
h_p=-\frac{1}{2}(\gamma \tilde{B}-\int dp' V(p-p')\bar{\sigma}(t,x,p'),
\end{eqnarray*}
$V$ is the inter-particle potential, ${\tilde{B}}$ an external magnetic field, and $\gamma$ is the gyromagnetic ratio. Open mathematical problems of interest for physics concern existence, regularity, validation, the relaxation times for spin-diffusion, and time asymptotic behaviour in general.\\
\hspace{1cm}\\
In this paper $p_1, p_2$ and $p'_1, p'_2$ denote post-collisional and pre-collisional moments in an elastic collision,
\begin{eqnarray*}
p_1=\frac{p'_1+p'_2}{2}+\frac{|p'_1-p'_2|}{2}\omega,\\
p_2=\frac{p'_1+p'_2}{2}-\frac{|p'_1-p'_2|}{2}\omega,
\end{eqnarray*}
where $\omega\in\mathcal{S}^2$ and $p_1,p_2,p'_1,p'_2\in \mathbb{R}^3$.
The distribution function $\rho$ with values in ${\mathcal H}_2(\mathbb C)$ is assumed to evolve according to the kinetic equation
\begin{eqnarray}
D\rho:=\frac{\partial}{\partial t} \rho + p\cdot \bigtriangledown_x\rho=Q(\rho)
\end{eqnarray}
with the { \it collision term} of [JM],
\begin{eqnarray}
Q(\rho)=
\int_{\mathbb{R}^3\times\mathcal{S}^2} dp_2d\omega B(p_2-p_1,\omega)
(\{[\tilde{\rho}_1,\rho_{1'}]_+\Tr(\tilde{\rho}_2\rho_{2'}) \nonumber\\
-[\rho_1,\tilde{\rho}_{1'}]_+\Tr(\rho_2\tilde{\rho}_{2'})\}-\{[{\rho}_{1'}\tilde{\rho}_2\rho_{2'},\tilde{\rho}_1]_+
-[\tilde{\rho}_{1'}\rho_2\tilde{\rho}_{2'},\rho_1]_+\}).
\end{eqnarray}
Here $\tilde{\rho}=I-\rho$, and $[.,.]_+$ denotes an anti-commutator.
This collision term with kernel $B=1$, also appears as the dissipative collision term
in the Hubbard-Boltzmann model, which describes the evolution of interacting spin-$\frac{1}{2}$ fermions on a lattice (see [LMS]).
In the present paper the collision kernel $B(z,\omega)$ is assumed to satisfy\\
i) $0\leq B\in L^1(\mathcal{B}_R\times\mathcal{S}^2):=L^1(\{z\in\mathbb{R}^3\};|z|\leq R\}\times\mathcal{S}^2)$ for $R>0$, ii) $B(z,\omega)=B(|z|,|(z,\omega)|)$,\\
iii) $(1+|z|^2)^{-1}\int_{z+\mathcal{B}_R}dv\int_{\mathcal{S}^2}B(v,\omega)d\omega\rightarrow 0,$ when $|z|\rightarrow \infty,$ $R\in (0,\infty)$.\\
Obviously the two terms in the integrand of (1.4) with a trace factor, are hermitean. For the remaining two terms that property follows by a change of variables $\omega\rightarrow-\omega$.
Separating the I- and $\rho$-part of $\tilde{\rho}$, we notice that
the ensuing terms in (1.4) without any I-factor, formally cancel each others.\\
The main result of the paper is the following global existence theorem for the initial value problem of (1.3).
\begin{theorem}
Consider the equation (1.3) with initial value $\rho_0=\frac{1}{2}(f_0I+\bar{\sigma}_0\sigma)$. Suppose that $f_0\in L^{\infty}\cap L^1([0,1]^3\times \mathbb {R}^3)$ resp. $\bar{\sigma}\in(L^{\infty}\cap L^1([0,1]^3\times \mathbb {R}^3))^3$ , with $\int(1+p^2)f_0(x,p)dxdp<\infty$,
$0\leq f_0\leq 2$ and $\bar{\sigma}_0^2\leq \min(f_0^2,(2-f_0)^2)$. Then the equation (1.3) with initial value $\rho_0$, has a bounded integrable solution $\rho=\frac{1}{2}(fI+\bar{\sigma}\sigma)$ for $t>0$ with $0\leq f\leq 2$, and $\bar{\sigma}^2\leq \min(f^2,(2-f)^2)$ . The number density $f$ conserves mass and first moments, and has the kinetic energy bounded by its initial value.
\end{theorem}
\hspace{1cm}\\
The extension of this result to the more general and physically important equation (1.2) remains open.
Linearized versions of (1.3) are also discussed in [JM], but again we are not aware of any related mathematical studies.
However, phenomenologically modelled, {\it simplified linear} Boltzmann equations with spin, introduced in spintronics for semiconductor hetero-structures, such as the equation
\begin{eqnarray}
\frac{\partial}{\partial t} \rho + v\cdot \bigtriangledown_x\rho
+E\cdot  \bigtriangledown_v\rho =Q(\rho) + Q_{SO}(\rho)+Q_{SF}(\rho),
\end{eqnarray}
have been analyzed mathematically. Here $E$ is an electric field, and $Q$ is the collision operator for collisions without spin-reversal in the linear BGK approximation
\begin{eqnarray*}
 \int _{\mathbb{R}^3} \alpha(v,v')(M(v)\rho(v')-M(v')\rho(v))dv',
\end{eqnarray*}
with $M$ denoting a normalized Maxwellian. The spin-orbit coupling generates an effective field $\Omega$ making the spins precess. The corresponding spin-orbit interaction term $Q_{SO}(\rho)$ is given by $\frac{i}{2}[\Omega\cdot \sigma, \rho]$. Finally $Q_{SF}(\rho)$ is a spin-flip collision operator, in relaxation time approximation given by
\begin{eqnarray*}
Q_{SF}(\rho)= \frac{{\rm tr} \rho I_2-2 \rho}{\tau_{sf}},
\end{eqnarray*}
with $\tau _{sf}>0$ the spin relaxation time.
Mathematical properties of the full equation (1.5), such as existence, uniqueness, and asymptotic behaviour, have been studied
in particular by the French group around Ben Abdallah with coworkers and students (see [EH1, EH2] and references therein).\\

\section{Proof of Theorem 1.1.}
The approach from [D] and [PLL] for scalar Fermi-Dirac type Boltzmann equations, is here adapted to the initial value problem of the spin-dependent equation (1.3) to prove Theorem 1.1. Let $B_j$ be the restriction of $\min(B,j)$  to the set $p_1^2+p_2^2\leq j^2$. We first consider the approximation using the truncated kernel $B_j$ in the collision operator $Q$, and denote the corresponding collision operator by $Q_j$.
For $\rho=\frac{1}{2}(fI+\bar{\sigma}\sigma)$, set
\begin{eqnarray*}
F(t,x,p)=f(t,x,p) \hspace{.5cm} {\rm for} \hspace{.5cm} 0\leq f\leq 2, \hspace{.5cm} =0 \hspace{.5cm} {\rm for} \hspace{.5cm} f<0, \hspace{.5cm}
=2 \hspace{.5cm} {\rm for} \hspace{.5cm} f>2 \\
\Sigma(t,x,p)=\bar{\sigma}(t,x,p)\hspace{.5cm} {\rm when} \hspace{.5cm}  \min(F^2,(2-F)^2)\geq \bar{\sigma}\cdot\bar{\sigma},\\
 {\rm else} \hspace{.5cm}\Sigma(t,x,p)=
\frac{\min(F,2-F)\bar{\sigma}}{\sqrt{\bar{\sigma}\cdot\bar{\sigma}}}(t,x,p).
\end{eqnarray*}
Then ${R}:=\frac{1}{2}(FI+\Sigma\sigma)\geq 0$, and $(I-R)\geq 0$.
Consider the equation
\setcounter{equation}{0}
\begin{eqnarray}
D\rho=Q_j({R})
\end{eqnarray}
for $t>0$ with initial value $\rho_0$, where $\rho_0=\frac{1}{2}(f_0I+\bar{\sigma_0}\sigma)$ and $(f_0,\bar{\sigma}_0)\in (L^{\infty}([0,1]^3\times\mathbb{R}^3))^4$ .
\setcounter{theorem}{0}
\begin{lemma}
The equation (2.1) for the truncated kernel $B_j$ with initial value $\rho_0$, has locally in time a unique hermitean $L^{\infty}$-solution $\rho=\frac{1}{2}(fI+\bar{\sigma}\sigma)$.
\end{lemma}
\underline {Proof} Set
\begin{eqnarray}
T_j\rho^{\#}(t,x,p):=T_j\rho(t,x+tp,p)=\rho_0(x,p)+\int_0^t Q_j(R)(s,x+sp,p)ds.
\end{eqnarray}
Obviously $\rho$ is a solution of the the initial value problem for (2.1) with collision operator $Q_j$, if and only if it is a fixed point of $T_j$. So it is enough to prove that the operator $T_j$ is contracting in $L^{\infty}$ when $0\leq t\leq t_0$ for some $t_0>0$ and small enough.\\
Now
\begin{eqnarray*}\\
T_j(\rho^1)-T_j(\rho^2)= \int^t ds\int dp_2d\omega B_j(p_2-p_1,\omega)
(\{[\tilde{R}^1_1,R^1_{1'}]_+\Tr(\tilde{R}^1_2R^1_{2'}) \nonumber\\
-[R^1_1,\tilde{R}^1_{1'}]_+\Tr(R^1_2\tilde{R}^1_{2'})\}
-\{[{R}^1_{1'}\tilde{R}^1_2R^1_{2'},\tilde{R}^1_1]_+
-[\tilde{R}^1_{1'}R^1_2\tilde{R}^1_{2'},R^1_1]_+\})\\
-\int^t ds\int dp_2d\omega B_j(p_2-p_1,\omega)
(\{[\tilde{R}^2_1,R^2_{1'}]_+\Tr(\tilde{R}^2_2R^2_{2'})
-[R^1_2,\tilde{R}^2_{1'}]_+\Tr(R^2_2\tilde{R}^2_{2'})\}\nonumber\\
-\{[{R}^2_{1'}\tilde{R}^2_2R^2_{2'},\tilde{R}^2_1]_+
-[\tilde{R}^2_{1'}R^2_2\tilde{R}^2_{2'},R^2_1]_+\}).
\hspace{1cm}\\
\end{eqnarray*}
Assume that $\rho^{1\#}$ and $\rho^{2\#}$ are continuous in $t$  with respect to the norm
\begin{eqnarray*}
 \parallel \rho(t,.)\parallel_{\infty}:=
\sup _{(x,p)\in\mathbb{T}^3\times \mathbb{R}^3}\parallel \rho(t,x,p)\parallel.
 \end{eqnarray*}
 After some computations and using the bounds on $F^1,F^2,\Sigma^1,\Sigma^2$, one obtains
\begin{eqnarray*}
\parallel T_j(\rho^1)(t)-T_j(\rho^2)(t)\parallel_{\infty}\leq
tC_j\sup_{0\leq s\leq t}\parallel R^1(s)-R^2(s) \parallel_{{\infty}}.
\end{eqnarray*}
Here $C_j$ denotes a generic constant. But it is easy to see from its definition, that the mapping $\rho\rightarrow R$ is Lipschitz continuous in the $\parallel.\parallel_{\infty}$-norm, and so
\begin{eqnarray*}
\parallel T_j(\rho^1)(t)-T_j(\rho^2)(t)\parallel_{\infty}\leq t C_j\sup_{0\leq s\leq t}\parallel \rho^1(s)-\rho^2(s) \parallel_{{\infty}}.
\end{eqnarray*}
Hence $T_j$ is contracting  on $[0,t_0]$ for $t_0>0$ and sufficiently small, with $t_0$ independent of $\rho_0$. Moreover, $T_j$ preserves the hermitean property, hence the solution $\rho$ is hermitean. $\square$\\
\hspace{1cm}\\
It follows that equation (1.3) for $Q=Q_j$ holds locally in time if $\rho=R$, which is equivalent to $\rho\geq0$ and $I-\rho\geq0$.
The collision term (1.4) coincides after a change of variables with the collision term $\mathcal{C}_{diss}$ of [LMS], where a splitting into gain and loss term is introduced.
The gain term
\begin{eqnarray*}
\mathcal{G}_j(\rho)(p_1)=\int dp_2d\omega B_j(p_2-p_1,\omega)
(\rho_{1'}\Tr(\tilde{\rho}_2\rho_{2'})
+\Tr(\rho{_2'}\tilde{\rho}_{2})\rho_{1'}
-{\rho}_{1'}\tilde{\rho}_2\rho_{2'}
-{\rho}_{2'}\tilde{\rho}_{2}\rho_{1'}).
\end{eqnarray*}
is hermitean, if $\rho$ is hermitean. It holds
\begin{lemma}
{\rm [LMS]} If $\rho\geq0$ and $I-\rho\geq0$, then $\mathcal{G}_j(\rho)\geq 0$.
\end{lemma}
\underline{Idea of proof}. The result follows from the matrix inequalities
\begin{eqnarray*}
A({\Tr} (BC) -BC)+C({\Tr}(BA)-BA)\geq0,\quad ({\Tr}(AB)-AB)C+({\Tr}(CB)-CB)A\geq0
\end{eqnarray*}
for hermitean $n\times n$ matrices $A,B,C\geq0$ introduced into this area by [LMS], and using the change of variables $\omega\rightarrow-\omega$. $\square$\\
\\
\hspace{1cm}\\
With
\begin{eqnarray*}
\mathcal{D}_j(\rho)(p_1)=\int dp_2d\omega B_j(p_2-p_1,\omega)
(\Tr(\rho_{2'}\tilde{\rho}_2)\rho_{1'}
+\Tr(\tilde{\rho}_{2'}{\rho_2})\tilde{\rho_{1'}}
-{\rho}_{2'}\tilde{\rho_{2}}{\rho}_{1'}
-\tilde{{\rho}_{2'}}{\rho}_{2}\tilde{\rho_{1'}}),
\end{eqnarray*}
the loss term becomes
\begin{eqnarray*}
\mathcal{D}_j(\rho)(p_1)\rho(p_1)+\rho(p_1)\mathcal{D}_j(\rho)(p_1).
\end{eqnarray*}
The loss term is hermitean together with $\mathcal{D}_j$. The collision operator $Q_j$ can now be written
\begin{eqnarray*}
Q_j(\rho)(p_1)=\mathcal{G}_j(\rho)(p_1)-\mathcal{D}_j(\rho)(p_1)\rho(p_1)-
\rho(p_1)\mathcal{D}_j(\rho)(p_1),
\end{eqnarray*}
and (2.1) becomes
\begin{eqnarray}
D_t\rho^{\#}+\mathcal{D}_j(R)^{\#}R^{\#}+R^{\#}\mathcal{D}_j(R)^{\#}=\mathcal{G}_j(R)^{\#}.
\end{eqnarray}
We next consider the special initial data $\rho_0$ such that for some $\eta_j>0$ and for all $|p|<j$, uniformly in $x$
\begin{eqnarray}
0<\eta_j\leq f_0\leq 2-\eta_j, \quad {\rm and} \quad \bar{\sigma}^2+\eta_j^2\leq \min(f^2_0,(2-f_0)^2).
\end{eqnarray}
Using the bounds on the norm of $T_j$, it follows that on a (short and j-dependent) positive time-interval $[0,t_j)$ which can be taken maximal in $[0,t_0]$, the solution satisfies (2.4) with a decreasing time-dependent $\eta_j(t)>0$.
Let $U(t,x,p)$ be the solution to $\partial_tU(t)=U(t)\mathcal{D}_j(\rho)^{\dagger}$, $0\leq t\leq t_j$, $U(0,x,p)=I$. The inverse $U^{-1}$, which exists initially by continuity since $U(0)=I$, satisfies the equation $\partial_tU^{-1}=-\mathcal{D}_j(\rho)^{\#}U^{-1}$. On $[0,t_j]$, the latter equation is solvable with the solution being the inverse of $U$. Hence (2.3) can be written
\begin{eqnarray*}
D_t(U(t)\rho^{\#}(t)U^{\dagger}(t))=
U(t)D_t\rho^{\#}(t) U^{\dagger}(t)+U(t)\mathcal{D}_j(\rho)^{\#}(t)\rho^{\#}(t)U^{\dagger}(t)\nonumber\\
+U(t)\rho^{\#}(t)\mathcal{D}_j(\rho)^{\#}U^{\dagger}(t)
=U(t)\mathcal{G}_j(\rho)^{\#}(t)U^{\dagger}(t).
\end{eqnarray*}
This gives
\begin{eqnarray*}
\rho^{\#}(t_j)=U^{-1}(t_j)\rho(0)U^{\dagger-1}(t_j)+\int _0^{t_j}dsU^{-1}(t_j)U(s)\mathcal{G}_j(\rho)^{\#}(s)U^{\dagger}(s)U^{\dagger-1}(t_j).
\end{eqnarray*}
The integral term is positive since $\mathcal{G}_j(\rho)$ is positive, and the term $U^{-1}(t_j)\rho(0)U^{\dagger-1}(t_j)$ satisfies (2.4) for some $\eta>0$. It follows that $t_j=t_0$ and (2.4) holds at $t_0$ for some $\eta>0$. An analogous reasoning holds for $I-\rho$. Global existence with positivity of $\rho$ and $I-\rho$ follow by iterating the arguments. Approximating by the above type of uniformly positive initial values an arbitrary initial value $\rho_0$ having $\rho_0\geq 0$ and $I-\rho_0\geq 0$, the existence result of Theorem 1.1 for $Q=Q_j$ follows by continuity.\\
 To prove the conservation properties for the number density $f^j$, consider the equation for $f^j$ resulting from (1.3) with $B=B_j$ (and dropping the $j$ from $f^j,\bar{\sigma}^j$)
\begin{eqnarray}
Df= Q_n(f,\bar{\sigma}),
\end{eqnarray}
where the number density part $Q_n$ of the collision operator is
\begin{eqnarray*}
Q_n(f_1,\bar{\sigma}_1)=\frac{1}{2} \int dp_2d\omega B_j(p_2-p_1,\omega)
\Big(([f_{1'}-\frac{1}{2}(f_1f_{1'}+\bar{\sigma}_1\cdot\bar{\sigma}_{1'})][f_{2'}
-\frac{1}{2}(f_2f_{2'}+\bar{\sigma}_2\cdot\bar{\sigma}_{2'})]\\
-[f_1-\frac{1}{2}(f_1f_{1'}+\bar{\sigma}_1\cdot\bar{\sigma}_{1'})]
[f_2-\frac{1}{2}(f_2f_{2'}+\bar{\sigma}_2\cdot\bar{\sigma}_{2'})])\\
-([\bar{\sigma}_{1'}-\frac{1}{2}(f_{1'}\bar{\sigma}_1+f_1\bar{\sigma}_{1'})]
\cdot[\bar{\sigma}_{2'}
-\frac{1}{2}(f_{2'}\bar{\sigma}_2+f_2\bar{\sigma}_{2'})]\\
-[\bar{\sigma}_1-
\frac{1}{2}(f_{1'}\bar{\sigma}_1+f_1\bar{\sigma}_{1'})]\cdot[\bar{\sigma}_2-
\frac{1}{2}(f_{2'}\bar{\sigma}_2+f_2\bar{\sigma}_{2'})])\Big).
\end{eqnarray*}
Conservation of mass, first moments, and kinetic energy follow by the usual change of variables argument.
\[\]
\underline{End of proof of Theorem 1.1.} It remains to prove that the initial value problem for (1.3)
has a solution $\rho$ also in the limit $j\rightarrow \infty$, i.e. for $B$ with the full domain $\mathbb{R}^3$. Extracting subsequences, let $(f,\bar{\sigma})$ denote the weak $L^1$-limit of $(f^j,\bar{\sigma}^j)$. Mass and first moments of $f$ are conserved, and its kinetic energy is bounded by the initial value. The weak limit $\rho$ defined by $(f,\bar{\sigma})$, satisfies (1.3). That can now be proved by the type of weak compactness arguments for $(\rho^j)_{j\in\mathbb{N}}$ that were introduced for the scalar Fermi-Dirac case of the Boltzmann equation by [PLL].
In fact, that proof holds step by step, when applied not just to the  non-cancelling terms of
\begin{eqnarray*}
f^j_{1'}f^j_{2'}(2-f^j_1)(2-f^j_2)- f^j_1f^j_2(2-f^j_{1'})(2-f^j_{2'})
\end{eqnarray*}
for the number density $f^j$ (i.e. those terms
with two or three factors), but to each (two-fold and triple) combination of number density $f^j$ and of
spin-components $\bar{\sigma}^j_k$, $k=1,2,3$ that appears.
Averaging lemma arguments and weak convergence steps can here be applied separately to each of the four scalar component equations of (1.3), by using that the number density $f^j$ satisfies $0\leq f^j\leq 2$, and the spins $\bar{\sigma}^j=( \bar{ \sigma}^j_1,\bar{\sigma}^j_2,\bar{\sigma}^j_3)$ satisfy $|\bar{\sigma}^j_k|\leq\min (f^j,2-f^j)$. The weak limit (extracting subsequences) in the collision term, of each such product of $f^j$´s and $\bar{\sigma}^j_k$´s, proves in this way to equal the corresponding product of elements of the limits $f$ and $\bar{\sigma}_k$´s.\\
\hspace{1cm}\\
In all such proofs, only an entropy dissipation argument in [PLL] to control various properties/estimates for $Q^+(f)=\int dp_2\int d\omega B(p_1-p_2,\omega) f_{1'}f_{2'}(1-f_1)(1-f_2)$ in terms of the corresponding ones for $Q^-(f)=\int dp_2\int d\omega B(p_1-p_2,\omega) f_1f_2(1-f_{1'})(1-f_{2'})$, has to be replaced in the present case by e.g. direct control.
The following example shows how that may be done. The $L^1$-convergence of a sequence $Q^+_j(f)$ is in [(134) of PLL]
carried out in terms of an already known convergence of the corresponding sequence $Q^-_j(f)$. That may be replaced by the following direct proof. Let $K$ be a fixed compact set in $\mathbb{R}^3_{p}$. For $p_1\in K$ and $j$ large, $\max(|p'_1|,|p'_2|)> \frac{|p_2|}{2})$ when $p_1^2+p_2^2>j$. Here the $B_j$'s are the earlier truncated versions of $B$. In the integral $Q^+(f)(p_1)=\int dp_2\int d\omega B(p_1-p_2,\omega) f'_1f'_2(1-f_1)(1-f_2)$
of [PLL], make a change of variable from $p_2$ to the one of $p'_1$ and $p'_2$ giving  $\max(|p'_1|,|p'_2|)$,  and estimate the other factors  in $f$ and $1-f$ by one. With $\omega=(\theta,\varphi)$, in these changes of variable the corresponding $\cos\theta$ or $\sin\theta$ is bounded away from zero. With $A_j(p_1,p_2)=\int {B_j(p_1,p_2,\omega)}d\omega$ and using the bounds on mass and energy, this gives
\begin{eqnarray}
\int dx\int_{K} dp_1|Q^+-Q^+_j|=\int dx\int_K dp_1dp_2d\omega (B-B_j)(p_1-p_2,\omega)f_1'f'_2(1-f_1)(1-f_2)\nonumber\\
 \leq  C\int dx\int dp_2f_2\int_K dp_1 (A-A_j)(p_1-p_2)dp_1\rightarrow 0,\quad \quad j\rightarrow \infty.
\end{eqnarray}
Replacing in our present case, one or more $f$'s by $\bar{\sigma}_k$'s with $|\bar{\sigma}_k|\leq f$, does not change the argument, nor does the introduction of further truncations of $B$.\\
In the [PLL] proof, the entropy dissipation argument also appears once before the case discussed in (2.6), namely for a weak $L^1$-convergence of $ Q^{\pm}(f^j)$ (extracting subsequences if necessary). There $Q^-(f^j)$ is bounded from above by the usual Boltzmann collision frequency which converges with $j$, weakly in $L^1$ on compacts. Then for $Q^+(f^j)£$, instead of comparing $Q^+$ with $Q^-$ as in [PLL], consider in our case directly integrals $\int_K dp_1\psi(p_1)Q^+(f^j)$ for test functions $\psi\in L^{\infty}$ with support in a compact $K\subset \mathbb{R}^3_p$, and argue similarly to (2.5) to prove a weak $L^1$-convergence of $Q^+(f^j)$ with respect to $K$. \\
\hspace{1cm}\\
Let $B_{\delta}$ denote a truncated version of $B$, regularized into $C_0^{\infty}$ and  zero for $|p|$, $|(p,\omega)|$, $|p|-|(p,\omega)|$ small and for $|p|$ large. When $B_{\delta}$ replaces the kernel $B$, weak convergence similarly holds for $Q^{\pm}(f^j)$ in $ L^q([0,1]^3\times\mathbb{R}^3_p)\times [0,T])$, $ 1\leq q<\infty $, $T>0$, and in $ L^{\infty}([0,1]^3\times\mathbb{R}^3_p\times(0,\infty))-* $. \\
After these preparations it is a routine but tedious work to reduce as in [PLL], the full proof to a proof of the following convergences in $L^q([0,1]^3\times\mathbb{R}^3\times(0,T))$ for all $1\leq q<\infty$, $T\in (0,\infty)$, while replacing his entropy dissipation argument by the above direct estimates. By the reduction it only remains to prove that
\begin{eqnarray}
Q^+_{\delta}(f^j,f^j)\rightarrow Q^+_{\delta}f,f),\quad Q^3_{\delta}(f^j,f^j)\rightarrow Q^3_{\delta}(f,f),\quad Q^4_{\delta}(f^j,f^j)\rightarrow Q^4_{\delta}(f,f),\quad j\rightarrow \infty,
\end{eqnarray}
and for $\psi\in C_0^{\infty}(\mathbb{R}^3)$, the convergence in $L^q([0,1]^3\times(0,T),$ $1\leq q<\infty$ of
\begin{eqnarray}
\int_{\mathbb{R}^3}\psi dp\int_{\mathbb{R}^3} dp_*\int_{\mathcal{S}^2} d\omega B_{\delta}{f^j}'{f^j}'_*f^j_*\rightarrow \int_{\mathbb{R}^3}\psi dp\int_{\mathbb{R}^3} dp_*\int_{\mathcal{S}^2} d\omega B_{\delta}f'f'_*f_*\quad j\rightarrow \infty.
\end{eqnarray}
Here $ Q^{\pm}_{\delta} $ is the usual quadratic gain term of the Boltzmann equation, but with as kernel the truncated version $B_{\delta}$, and
\begin{eqnarray*}
Q^3_{\delta}(f,f)=\int_{\mathbb{R}^3} dp_*\int_{\mathcal{S}^2} d\omega B_{\delta}f_*f',     \quad
Q^4_{\delta}(f,f)=\int_{\mathbb{R}^3} dp_*\int_{\mathcal{S}^2} d\omega B_{\delta} f'_*f_*.
\end{eqnarray*}
For each combination of $f$'s and $\bar{\sigma}_k$'s, the study of (2.7-8) can again be carried out step by step exactly as in [PLL], a simple example to illustrate being the following quadratic gain term type integral with test function $\psi\in C_0^{\infty}(\mathbb{R}^3)$
\begin{eqnarray*}
\int dp_1dp_2d\omega B_{\delta} \bar{\sigma}^j_1(t,x,p´_1)\bar{\sigma}^j_2(t,x,p´_2)\psi(p_1).
\end{eqnarray*}
 It converges by an averaging argument to
 \begin{eqnarray*}
\int dp_1dp_2d\omega B_{\delta} \bar{\sigma}_1(t,x,p´_1)\bar{\sigma}_2(t,x,p´_2)\psi(p_1).
\end{eqnarray*}
Similarly working through all the many cases of (2.7-8), completes the proof of Theorem 1.1.
$\square$

\end{document}